\documentclass[lettersize,journal]{IEEEtran}
\usepackage{amsmath,amsfonts}
\usepackage{algorithmic}
\usepackage{algorithm}
\usepackage{array}
\usepackage[caption=false,font=normalsize,labelfont=sf,textfont=sf]{subfig}
\usepackage{textcomp}
\usepackage{stfloats}
\usepackage{url}
\usepackage{verbatim}
\usepackage{graphicx}
\usepackage{cite}
\usepackage{siunitx}
\usepackage{upgreek}
\usepackage{hyperref}
\usepackage{wrapfig}
\begin{document}

\title{A three-dimensional polarization-insensitive grating coupler tailored for 3D nanoprinting}

\author{Oliver Kuster, Yannick Augenstein, Carsten Rockstuhl, and Thomas Jebb Sturges

\thanks{

Oliver Kuster and Carsten Rockstuhl acknowledge support by the Deutsche Forschungsgemeinschaft (DFG, German Research Foundation) under Germany's Excellence Strategy via the Excellence Cluster 3D Matter Made to Order (EXC-2082/2-390761711) and from the Carl Zeiss Foundation via the CZF-Focus@HEiKA Program.

(\textit{Corresponding author: Oliver Kuster})

Oliver Kuster is with the Institute of Theoretical Solid State Physics, Karlsruhe Institute of Technology, 76131 Karlsruhe, Germany (email: oliver.kuster@kit.edu).

Yannick Augenstein is with Flexcompute, Boston, Massachusetts.

Carsten Rockstuhl is with the Institute of Theoretical Solid State Physics, Karlsruhe Institute of Technology, 76131 Karlsruhe, Germany; also with the Institute of Nanotechnology, Karlsruhe Institute of Technology, 76131 Karlsruhe, Germany.

Thomas Jebb Sturges is with the Institute of Nanotechnology, Karlsruhe Institute of Technology, 76131 Karlsruhe, Germany  (e-mail: thomas.sturges@kit.edu).

}}

\markboth{Journal of Selected Topics in Quantum Electronics}%
{Shell \MakeLowercase{\textit{et al.}}: A Sample Article Using IEEEtran.cls for IEEE Journals}


\maketitle


\begin{abstract}

Efficiently coupling light from optical fibers into photonic integrated circuits is a key step toward practical photonic devices.  
While a notable coupling can be achieved by out-of-plane couplers such as grating couplers, their basic planar geometry typically tends to be sensitive to the polarization of light. This is partly due to the fact that the design spaces of such grating structures—typically fabricated with techniques such as electron-beam lithography—are only two-dimensional with a simple extrusion into the vertical dimension.
This makes it challenging to optimize for both polarizations simultaneously, as performance typically degrades when trying to achieve high efficiency in both. As a result, conventional approaches either suffer from increased losses or require additional filtering components to account for different polarizations.
In this work, we present a fully three-dimensional, polarization-insensitive grating coupler which has a highly efficient simulated coupling efficiency of over $\mathbf{80\%}$ in both polarizations. This performance matches that of state-of-the-art couplers that are performant for one polarization only. This comes at the cost of a moderately larger size due to the lower refractive index materials typically available in 3D nanoprinting.
Our design method uses density-based topology optimization with a multi-objective approach that combines electromagnetic simulations with a fictitious heat-conduction model acting as a soft constraint to promote structural integrity. This ensures that the designed structures are feasible for fabrication.
Our work opens new possibilities for robust 3D photonic devices, enabling advanced integration, fabrication, and applications across next-generation photonics and electronics.


\end{abstract}


\section{Introduction}
3D nanoprinting enables the fabrication of nanophotonic devices on the scale of a few micrometers up to the centimeter scale.
The printing process makes it possible for us to consider free-form design in all three dimensions with voxel sizes smaller than the wavelength of the light.
Due to its accessibility and cost-effectiveness, 3D nanoprinting is emerging as a viable alternative to the traditional lithography process of designing nanophotonic circuits. 
The ability to print photonic integrated circuit (PIC) components on demand in a fast and low-cost manner is particularly appealing compared to more expensive methods like electron-beam lithography.
While still an active front of research, significant progress has been made in 3D nanoprinting technologies recently.
More recent developments in the field include higher throughput, novel high refractive index materials, and even the ability to push the minimum feature size down to $\SI{100}{nm}$ \cite{Hahn:19, barnerkowollik_3d_2017, chen_recent_2025, kiefer_multi-photon_2024, yang_multi-material_2021, hahn_rapid_2020, bauer_sinterless_2023}. 
All of these developments enable efficient designs for PICs at even smaller length scales \cite{nagarajan_large-scale_2005, xiao_recent_2023, dietrich_situ_2018, huang2025multiphoton}. Moreover, and particularly important in the context of the special issue at hand, 3D nanoprinting offers a route towards three-dimensional photonic devices, significantly expanding the space available to implement functional devices. 

To build on these advances and to enable scalable and performant photonic computing devices, highly efficient basic components are required.
One such PIC component is a waveguide coupler, which couples light from an optical fiber into or out of the PIC.
The two most popular strategies for interfacing with PICs are edge couplers, which couple in-plane, and diffraction gratings, which are a vertical coupling strategy.
Both of these strategies can typically achieve a coupling efficiency of above $80\%$ \cite{mu2020edge, papes2016fiber, strauch2025inverse, vitali_highly_2022, hammond_multi-layer_2022, watanabePerpendicularGratingCoupler2017}, but grating couplers in particular tend to be sensitive to the polarization of the incoming light.
Typically, grating coupler designs involve only a few adjustable parameters, such as height, width, and pitch of the grating.
More advanced optimization schemes use topology optimization to reach higher coupling efficiencies in their design, but only consider a two-dimensional density-based design approach, where the final design is extruded into the third dimension.
Almost all of these designs are also tailored towards Silicon-on-Insulator (SoI) or Complementary Metal-Oxide-Semiconductor (CMOS) compatible platforms, which are mostly two-dimensional designs \cite{shao2010highly, chen2011polarization, alonso-ramos_polarization-independent_2012, song2015polarization, zhang2015polarization, xue_two-dimensional_2019, ma2020polarization, wang_polarization-independent_2021, zhou_polarization-independent_2022, zhang_polarization-insensitive_2022, nisar_polarization-insensitive_2022, kohli_c-_2023, sultan_highly_2025, cheng_ultra-compact_2025, wu_ultra-low-loss_2025}.

An overview of various reported designs for polarization-independent couplers is given in Tab.~\ref{tab:overview}. We note that the table shows the reported coupling efficiencies of the designs.
While the majority of the grating couplers presented in Tab.~\ref{tab:overview} also include experimental results, the reported experimental coupling efficiencies are lower than the simulated ones.
We also want to note that the "height" in the table refers to the design height, which typically coincides with the waveguide height and the etch depth.

Even though two-dimensional density-based design strategies are well-suited for more traditional lithography techniques such as electron-beam lithography, they are also limited in their accessible design space. Fully three-dimensional designs are rare \cite{trappen_3d-printed_2020, jung_ultra-broadband_nodate}.
This limited design space makes the design of efficient and flexible components especially challenging.
For example, attempts to design polarization-insensitive couplers using traditional two-dimensional techniques generally result in a compromise—neither polarization achieves the efficiency possible in a single-polarization-optimized device. This prompts the question: how much better can a fully three-dimensional design perform compared to a planar design?
\begin{table*}
\centering
\caption{Performance comparison of polarization-insensitive grating couplers in recent years (simulated results)}
    \begin{tabular}{c|c|c|c|c}
        Year &  Reference & Material \& Features & CE (sim. / exp.) & Dimension \\
        \hline
        2010 & Shao \textit{et al.} \cite{shao2010highly} & SOI uniform grating coupler with a T-shaped grating & $58\%$ / - & Height: $\SI{0.26}{\upmu m}$\\
        2011 & Chen \textit{et al.} \cite{chen2011polarization} & SOI nonuniform grating coupler with 2D grating features & $64\%$ / - & Height: $\SI{0.34}{\upmu m}$\\
        2012 & Ramos \textit{et al.} \cite{alonso-ramos_polarization-independent_2012}& SOI uniform grating coupler & $55\%$ / - & Height: $\SI{0.5}{\upmu m}$\\
        2015 & Song \textit{et al.} \cite{song2015polarization} & SOI nonuniform grating coupler & $20\%$ / $18\%$ & Height: $\SI{0.22}{\upmu m}$\\
        2015 & Zhang \textit{et al.} \cite{zhang2015polarization} & SOI uniform grating coupler with an added directional coupler & $74\%$ / - & Height: $\SI{0.46}{\upmu m}$\\
        2019 & Xue \textit{et al.} \cite{xue_two-dimensional_2019} & SOI grating coupler with elliptical-like etching pattern & $55\%$ / $38\%$ &$\SI{11}{\upmu m} \times \SI{11}{\upmu m} \times \SI{0.22}{\upmu m}$\\
        2020 & Ma \textit{et al.}\cite{ma2020polarization}& SOI uniform grating coupler & $51\%$ / - & Height: $\SI{0.3}{\upmu m}$\\
        2021 & Zhang \textit{et al.}\cite{zhang_high-efficiency_2021}& SOI 2D grating coupler with four access waveguides & $65\%$ / - & $\SI{5}{\upmu m} \times \SI{5}{\upmu m} \times \SI{0.22}{\upmu m}$\\
        2021 & Wang \textit{et al.}\cite{wang_polarization-independent_2021}& SOI uniform grating coupler optimized with genetic algorithms & $55\%$ / $18\%$ & Height: $\SI{0.22}{\upmu m}$\\
        2022 & Zhou \textit{et al.} \cite{zhou_polarization-independent_2022} & SOI nonuniform grating coupler & $44\%$ / $41\%$ & $\SI{12}{\upmu m} \times \SI{10}{\upmu m} \times \SI{0.22}{\upmu m}$\\
        2022 & Zhang \textit{et al.} \cite{zhang_polarization-insensitive_2022} & SOI grating coupler with rectuangular grating & $30\%$ / $23\%$ & $\SI{6}{\upmu m} \times \SI{20}{\upmu m} \times \SI{0.24}{\upmu m}$\\
        2022 & Nisar \textit{et al.} \cite{nisar_polarization-insensitive_2022} & SOI uniform rectuangular grating & $60\%$ / - & Height: $\SI{0.5}{\upmu m}$\\
        2023 & Kohli \textit{et al.} \cite{kohli_c-_2023} & SOI uniform grating coupler with bottom mirror & $91\%$ / $73\%$ & Height: $\SI{0.29}{\upmu m}$\\
        2025 & Sultan \textit{et al.} \cite{sultan_highly_2025} & SOI uniform grating coupler with bottom mirror & $89\%$ / - & $\SI{15}{\upmu m} \times \SI{8.5}{\upmu m} \times \SI{0.340}{\upmu m}$\\
        2025 & Cheng \textit{et al.} \cite{cheng_ultra-compact_2025} & SOI 2D topology optimized grating coupler & $52\%$ / - & $\SI{3.6}{\upmu m} \times \SI{3.6}{\upmu m} \times \SI{0.22}{\upmu m}$\\
        2025 & Wu \textit{et al.} \cite{wu_ultra-low-loss_2025} & SOI 2D topology optimized grating coupler with two access waveguides & $80\%$ / $79\%$ & $\SI{2.8}{\upmu m} \times \SI{2.8}{\upmu m} \times \SI{0.22}{\upmu m}$\\
        2025 & This work & 3D topology optimized grating coupler & $83\%$ / - & $\SI{18}{\upmu m} \times \SI{18}{\upmu m} \times \SI{4.5}{\upmu m}$\\
    \end{tabular}
    \label{tab:overview}
\end{table*}

\begin{figure*}[h!]
    \centering
    \includegraphics[width=\linewidth]{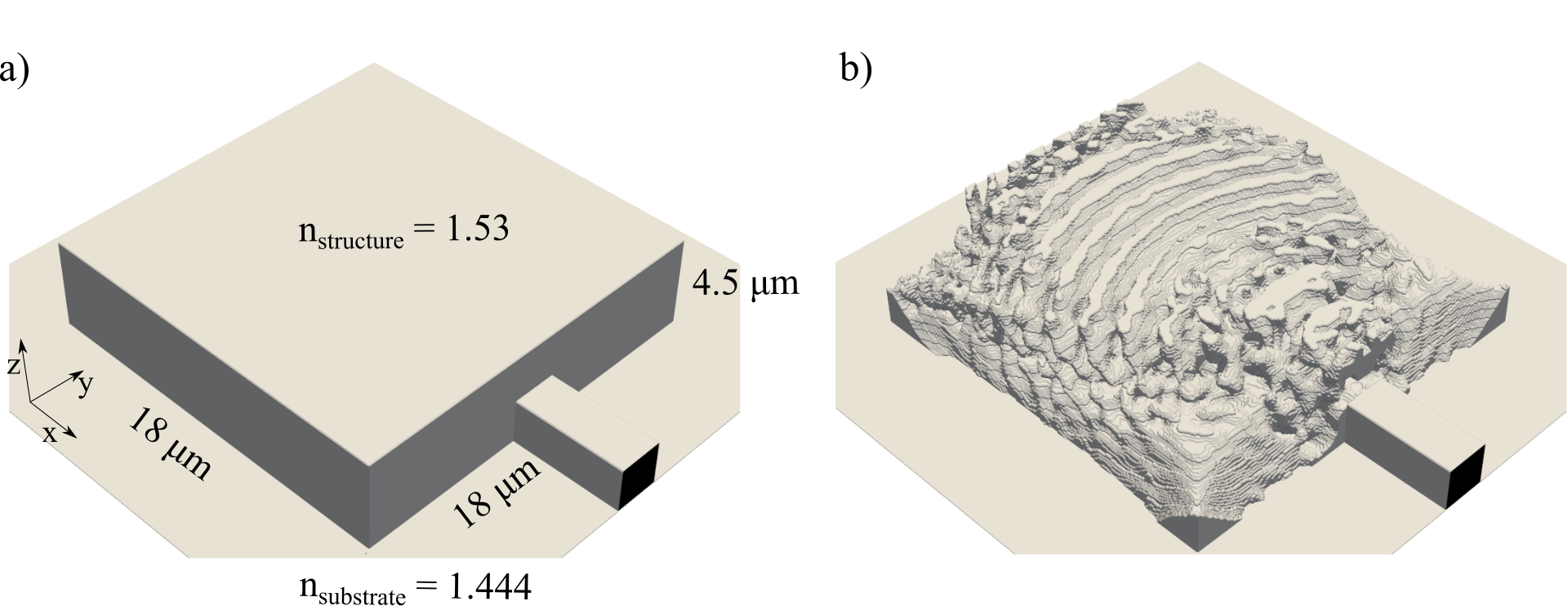}
    \caption{a) The design setup. Our design region has a size of $\qtyproduct{18 x 18 x 4.5}{\upmu m}$ and is illuminated from the top at an angle of $10^\circ$. A waveguide is placed at the edge of the design region. Both the design region and the waveguide have a refractive index of $n_\text{structure}=1.53$. They both sit on top of a substrate with a refractive index of $n_\text{substrate}=1.444$ and are embedded in air $n_\text{air}=1$. b) An optimized, structurally integral grating coupler design. This is a 3D render of the coupler, sitting on top of the substrate with the waveguide.}
    \label{fig:sketch_and_design}
\end{figure*}

3D additive manufacturing allows us to access a full three-dimensional design space, which means a larger parameter space for density-based approaches.
Instead of being limited to in-plane optimized structures, every voxel in the volume of interest can be used as a design parameter \cite{dana_free-standing_2024, eivgi_benchmarking_2025, liu_3d_2022, sturges_inverse-designed_2024, chen_anomalous_2024}.
By employing gradient-based optimization, in particular topology optimization, we can accommodate millions of degrees of freedom at the cost of only one additional simulation per forward simulation \cite{jensen_topology_2011, molesky_inverse_2018, sigmund_topology_2013}.
However, a full free-form design approach comes with problems of its own. The most optically performant free-form devices are usually not fabricable due to their lack of structural integrity or the presence of disconnected, floating features. Additionally, such designs may contain enclosed cavities that trap photoresist during fabrication, thereby degrading the optical performance.
To address these issues, we employ a virtual heat conduction model as an auxiliary objective during optimization. This acts as a soft constraint that encourages material connectivity and penalizes isolated or unsupported features \cite{cool_practical_2025, li_structural_2016, kuster_inverse_2025}.

In this work, we present a fully three-dimensional polarization-insensitive grating coupler of size $\qtyproduct{18 x 18 x 4.5}{\upmu m}$, optimized using topology optimization.
A sketch of the setup is shown in Fig.~\ref{fig:sketch_and_design} a), with the optimized design shown in Fig.~\ref{fig:sketch_and_design} b).
Our design can reach a simulated coupling efficiency of roughly $83\%$ independent of the polarization of the incoming light.

\section{Method}
Our system consists of a design region $D\subseteq\mathbb{R}^3$, a waveguide, a substrate, and a Gaussian source to approximate the fiber mode. The goal is to couple both the $x$- and $y$-polarized light from the fiber with a wavelength of $\lambda_0 = \SI{1.55}{\upmu m}$ as efficiently as possible into a waveguide of size $\SI{2}{\upmu m} \times \SI{2}{\upmu m}$. Namely, the $x$-polarized wave will be coupled into the fundamental transverse electric (TE$_{00}$) mode and the $y$-polarized light into the fundamental transverse magnetic (TM$_{00}$) mode, respectively.
The waveguide size is chosen so it can be coupled to the waveguides on a chip.
As we aim to design a device that can be printed using 3D nanoprinting, we assume a polymer resist with a refractive index $n_\text{structure}=1.53$ for both the design region and the waveguide. Our design is placed on top of a SiO$_2$ substrate with refractive index $n_\text{substrate}=1.44$.
We note that our choice of the refractive index of the polymer represents the refractive index of a typical polymer resist (e.g., IP-Dip for the Nanoscribe machine), but can be changed to any other refractive index value when optimizing for a different use case.
The waveguide is placed at the edge of the design region and possesses a cross section of $\qtyproduct{2 x 2}{\upmu m}$.
The entire setup is surrounded by air with $n_\text{air}=1$.
We approximate the output field of the fiber by a Gaussian profile with a mode-diameter of $\SI{10.4}{\upmu m}$ with its center at the center of the design region. We assume the Gaussian source to be slightly tilted at an angle of $\theta = 10^\circ$ from the $z$-axis, pointing towards the waveguide.

To solve Maxwell's equations, we use a finite-difference time-domain (FDTD) solver provided by Flexcompute's Tidy3D.
To ensure that the structure does not collapse on itself and is fully connected, we employ a virtual heat strategy \cite{cool_practical_2025, li_structural_2016 ,kuster_inverse_2025}.
The material is modeled as both a heat source and a good heat conductor, while void regions are modeled as having a poor heat conductivity.
By simulating the material as heat sources and designating the substrate as heat sinks, we can use the total heat of the system as a regularization term to promote connectivity of the material.
The total heat here is minimized as a sub-objective in addition to the sub-objective of the optical performance of the device.
Furthermore, we also need to ensure the void connectivity to avoid the formation of cavities inside the design. By separately employing the virtual heat strategy on the void as well, we promote the connection of the void regions to heat sinks placed at the interface between the design region and air. 
Together, these two virtual heat sub-objectives lead to a grating coupler design which can be directly printed using 3D nanoprinting techniques.
Our in-house heat solver uses a finite-element method, which solves the steady-state heat equation using eight-node hexahedral solid elements.
A more detailed explanation of all the methods used in this section can also be found in \cite{kuster_inverse_2025}.

The design procedure uses four separate simulations in the forward direction. We use two FDTD simulations, one for each $x$- and $y$-polarization, and two finite-element steady state heat equations, one for the material and one for the void.
These simulations give us the sub-objectives, which we then use to construct our multi-objective figure of merit (FoM).
Since we use gradient-based optimization, we need one additional simulation per forward simulation to calculate the gradients using the adjoint method in combination with automatic differentiation. This results in eight simulations total per iteration for the optimization.
The electromagnetic sub-objective is given by calculating the coupling efficiency of the incoming electromagnetic wave into the fundamental waveguide modes.
To do so, we extract the complex mode amplitudes inside the waveguide, which are given as $\alpha^+_{\text{TE}_{00}}$ for the fundamental TE mode and $\alpha^+_{\text{TM}_{00}}$ for the fundamental TM mode. Each mode is considered only in the forward propagation direction and normalized by the input power of the system. 
Our electromagnetic sub-objective for a single wavelength is then given as
\begin{equation}
    \mathcal{L}_\text{EM}(\rho) = 1 - \frac{|\alpha^+_{\text{TE}_{00}}|^2 + |\alpha^+_{\text{TM}_{00}}|^2}{2}\, ,
\end{equation}
since we want to maximize the transmission, but minimize our objective.
The sub-objectives for the heat simulations are defined as the total heat of the system, renormalized by a parameter $\tau \in \mathbb{R}$ that controls the strength of the connectivity constraint, as follows:

\begin{equation}
    \mathcal{L}_\text{heat}(\rho) = \frac{\sum_{x, y, z \in D}T(x, y, z) - \tau}{\tau}\, .
\end{equation}

The threshold $\tau$ determines the degree of connectivity. A higher value for $\tau$ means that $\mathcal{L}_\text{heat}(\rho)$ will be negative at higher values of $T$ than for a lower value of $\tau$.
We will elaborate on how we choose $\tau$ later.
Our problem is parametrized using a density-based approach, where $\rho(x, y, z) \in [0, 1]$ is the density of the material, while $1-\rho(x, y, z)$ is the density of the void at the point $(x, y, z)\in D$.
The design region itself consists of a grid of $300\times 300\times 75$~voxels, resulting in $6.75$ million degrees of freedom, which converts to an individual voxel size of $\SI{60}{nm}$. We note that we use mirror symmetry in the $y=0$-plane of the design, which reduces the effective size of our grid to $300\times 150\times 75$~voxels.
Note that an additional adaptive mesh refinement of $\SI{20}{px}$ per wavelength inside the medium is done to improve the accuracy and efficiency in the electromagnetic simulation.
By mapping $\rho$ to physical quantities, specifically the relative electric permittivity $\varepsilon$ for the electromagnetic simulation and the heat conductivity $\kappa$ for the heat simulation, we can retrieve the sub-objectives.
By changing the values of $\rho(x, y, z)$ at every point in the grid, we can then start to optimize our system to iteratively design our grating coupler.

\begin{figure}
    \centering
    \includegraphics[width=\linewidth]{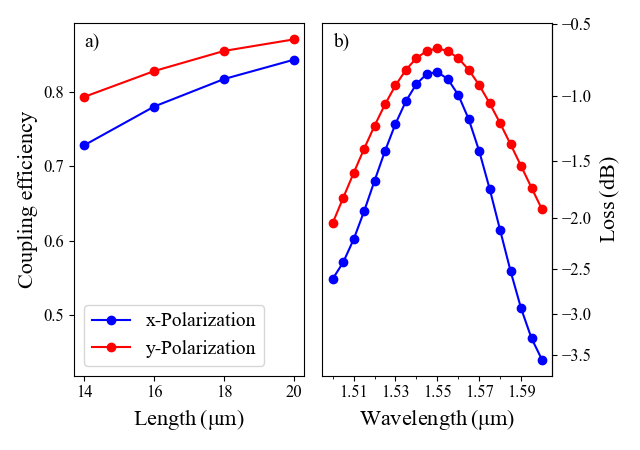}
    \caption{a) The coupling efficiency of both polarizations depends on the size of the design region. We look at four different sizes, which are given by $\qtyproduct{14 x 14 x 3.5}{\upmu m}$, $\qtyproduct{16 x 16 x 4}{\upmu m}$, $\qtyproduct{18 x 18 x 4.5}{\upmu m}$, and $\qtyproduct{20 x 20 x 5}{\upmu m}$, respectively. We note that while the design with size $\qtyproduct{20 x 20 x 5}{\upmu m}$ is the optically best performing one, free-floating artifacts start to appear. b) The coupling efficiency and loss for both polarization directions of the $\qtyproduct{18 x 18 x 4.5}{\upmu m}$ grating coupler as a function of the wavelength. We optimized for a single wavelength at $\SI{1.55}{\upmu m}$.
    Note that both figures share their y-axis.}
    \label{fig:loss_plot}
\end{figure}

We employ two additional filtering steps to account for fabrication constraints in our design. 
First, we use a conic filter provided by Tidy3D's adjoint module to enforce a minimum feature size in our problem. Then, we need to binarize our density. Since every step in the optimization needs to be done in a differentiable manner, we use an approximation of the Heaviside function
\begin{equation}
    f(x) = \frac{\tanh(\beta \cdot \alpha) + \tanh(\beta \cdot (x - \alpha))}{\tanh(\beta \cdot \alpha) + \tanh(\beta \cdot (1 - \alpha))}
\end{equation}
for binarization.
Here, $\alpha$ represents the center of the approximation, while $\beta$ defines the steepness of the function, corresponding to the degree of binarization. We choose $\alpha=0.5$ for all simulations.
We refer to the filtered and binarized density as $\hat{\rho} = \hat{\rho}(\rho)$, which are then mapped to the relative permittivity $\varepsilon(\hat{\rho}) \in [1, n_\text{structure}^2]$.
The thermal conductivity $\kappa(\rho) \in [10^{-5}, 1]$ is calculated directly from the density.

\begin{figure*}[h!]
    \centering
    \includegraphics[width=0.75\linewidth]{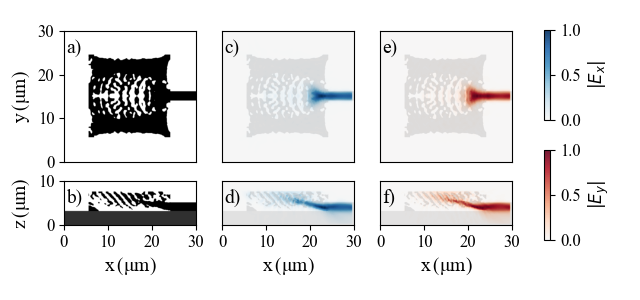}
    \caption{Various cross-sections of the optimized design. a) Cross-section of the design in the $x-y-$plane at the middle of the waveguide. b) Cross-section of the design in the $x-z-$plane at the middle of the design. c), d) The normalized $|E_x|$ field in frequency domain at a wavelength of $\SI{1.55}{\upmu m}$. e), f) The normalized $|E_y|$ field in frequency domain at a wavelength of $\SI{1.55}{\upmu m}$.}
    \label{fig:eps_field}
\end{figure*}

To balance the contributions of our sub-objectives, we use the softplus function
\begin{equation}
    \text{SP}(x) = \ln(1 + e^x)\, ,
\end{equation}
which is a differentiable approximation of a ramp function \cite{schubert2022inverse}. For negative values of $x$, the softplus function converges to $0$, allowing us to reduce the contributions from the heat sub-objectives once $x=0$ has been crossed.
We also choose $\tau$ so that we stop optimizing for connectivity once a certain level of connectivity is reached. The choice of the numerical value $\tau$ can be done by choosing a desired initial value $\mathcal{L}_\text{material/void}$ and deriving $\tau$ off of an initial heat simulation.
Finally, we can formulate our optimization problem as
\begin{align}
    \text{min }&\mathcal{L}(\rho) = \sqrt{\text{SP}\left(\mathcal{L}_\text{EM}\right)^2 + \text{SP}\left(\mathcal{L}_\text{material}\right)^2 + \text{SP}\left(\mathcal{L}_\text{void}\right)^2}\\
  &\text{s.t.}  \,\,\, (v_\text{ph}\nabla^2 - \frac{\partial^2}{\partial t^2})\mathbf{E}(t) = 0\,,\nonumber\\
  &\text{s.t.} \,\,\, -\kappa_m\nabla^2 u(x, y, z) = \rho\,,\nonumber\\
  &\text{s.t.} \,\,\, -\kappa_v\nabla^2 u(x, y, z) = 1-\rho\,,\nonumber\\
 &\text{s.t.} \,\,\, 0 \le\rho(x, y, z) \le 1\,,\nonumber
\end{align}
where $v_\text{ph}$ is the speed of light in the medium, $\mathbf{E}(t)$ is the electric field in time domain, $u(x, y, z)$ is the heat of the system, $\kappa_m$ the heat conductivity of the material, and $\kappa_v$ is the heat conductivity of the void. 

The design is parametrized with an initial distribution of $\rho(x, y, z)=0.5$ everywhere and then filtered at each step so that a minimum feature size of roughly $\SI{230}{nm}$ is enforced.
The density is then mapped to the electric permittivity and the heat conductivity for the respective simulations.
By enforcing either even or odd symmetry for our electromagnetic wave, we also get the correct symmetries for our fundamental waveguide modes inside the waveguide.

The gradients $\frac{\text{d}\mathcal{L}}{\text{d}\rho}$ are calculated by automatic differentiation and the adjoint method using Jax \cite{jax2018github} and Tidy3D's adjoint module.
Our optimization is done using the stochastic optimizer ADAM with a learning rate of $10^{-3}$ provided by optax \cite{deepmind2020jax} package. 
The full optimization runs for $200$ iterations. We increase the binarization $\beta$ after every $60$ steps from $100$ to $1000$ to $10000$ to gradually enforce the binarization, then simulate for another $20$ steps at the highest binarization to ensure convergence.

\section{Results}
In the following, we present the result of our optimizations. We re-optimize the device for different total volumes of the design region and compare the coupling efficiency for each polarization defined as the absolute-squared mode-amplitude for said polarization $|\alpha^+_\text{pol}|^2$.

We design grating couplers for four different sizes: $\qtyproduct{14 x 14 x 3.5}{\upmu m}$, $\qtyproduct{16 x 16 x 4}{\upmu m}$, $\qtyproduct{18 x 18 x 4.5}{\upmu m}$, and $\qtyproduct{20 x 20 x 5}{\upmu m}$, respectively.
For all four of these designs, we choose $\tau$ such that we initialize $\mathcal{L}_\text{material} = \mathcal{L}_\text{void} = 0.2$ to moderately enforce the connectivity constraints.
The coupling efficiency of each design is shown in Fig.~\ref{fig:loss_plot} a).

While increasing the design volume generally improves coupling efficiency, we observe that, unlike the remaining structures, the largest ($\qtyproduct{20 x 20 x 5}{\upmu m}$) design exhibits disconnected, floating features and lacks structural integrity. Although it achieves a high coupling efficiency of $85\%$, the resulting geometry is not physically realizable via 3D nanoprinting and can, therefore, be excluded as a viable design.
The disconnected feature is a free-floating part on the back side of the device. These features are small enough so that the optimizer no longer prioritizes connecting them properly.
We want to emphasize that, despite the free-floating parts being small, additional postprocessing will be required. Our method yields structures ready to print, and no additional postprocessing of the design is required.
The back side of the design region, in particular, seems prone to disconnected parts, with thinner structures extruding upward.
If the design region becomes too large, it is likely that these thin structures would become too thin and fall below the minimum feature size threshold, thus also becoming disconnected in the filtered and binarized density $\hat{\rho}$.
At this point, especially at higher binarizations $\beta$, the optimizer might struggle to properly connect the free floating part again. This is especially the case if the optimization can gain a bigger improvement in the figure of merit by increasing the coupling efficiency.
We show the heat distribution of the $\qtyproduct{20 x 20 x 5}{\upmu m}$ design in Fig.~\ref{fig:20}, where the disconnected region is clearly visible due to the higher heat.
We note that it is possible to find fully connected devices of this size (results not shown) by increasing the strength of the regularization of the connectivity constraint (i.e., decreasing $\tau$). However, the optical performance degrades to the point that it is worse than that of the next-smaller device investigated. Thus, it appears that the $\qtyproduct{18 x 18 x 4.5}{\upmu m}$ design is the best performing fabricable design.
An overview of all four designs analyzed in Fig.~\ref{fig:loss_plot} a) can be found in the supplementary material Fig.~\ref{fig:all-four}.


The $\qtyproduct{18 x 18 x 4.5}{\upmu m}$ design can be seen in Fig.~\ref{fig:sketch_and_design} b), with Fig.~\ref{fig:loss_plot} b) showing its wavelength-dependent coupling efficiency.

At the target wavelength of $\SI{1.55}{\upmu m}$, the design achieves a coupling efficiency of $82\%$ for the $x$-polarized source and a coupling efficiency of $84\%$ for the $y$-polarized source, averaging out to a coupling efficiency of $83\%$.
Our coupling efficiency for either polarization is comparable with values reported in literature for two dimensional grating couplers which were designed for a single polarization only\cite{huang_compact_2025, zhang_high-efficiency_2021, vitali_highly_2022, hammond_multi-layer_2022, watanabePerpendicularGratingCoupler2017}, and outperforms other polarization insensitive 2D grating couplers which have a performance range between $30\%$-$80\%$ without a bottom mirror \cite{shao2010highly, chen2011polarization, alonso-ramos_polarization-independent_2012, song2015polarization, zhang2015polarization, xue_two-dimensional_2019, ma2020polarization, wang_polarization-independent_2021, zhou_polarization-independent_2022, zhang_polarization-insensitive_2022, nisar_polarization-insensitive_2022, kohli_c-_2023, sultan_highly_2025, cheng_ultra-compact_2025, wu_ultra-low-loss_2025}.

\begin{figure}
    \centering
    \includegraphics[width=\linewidth]{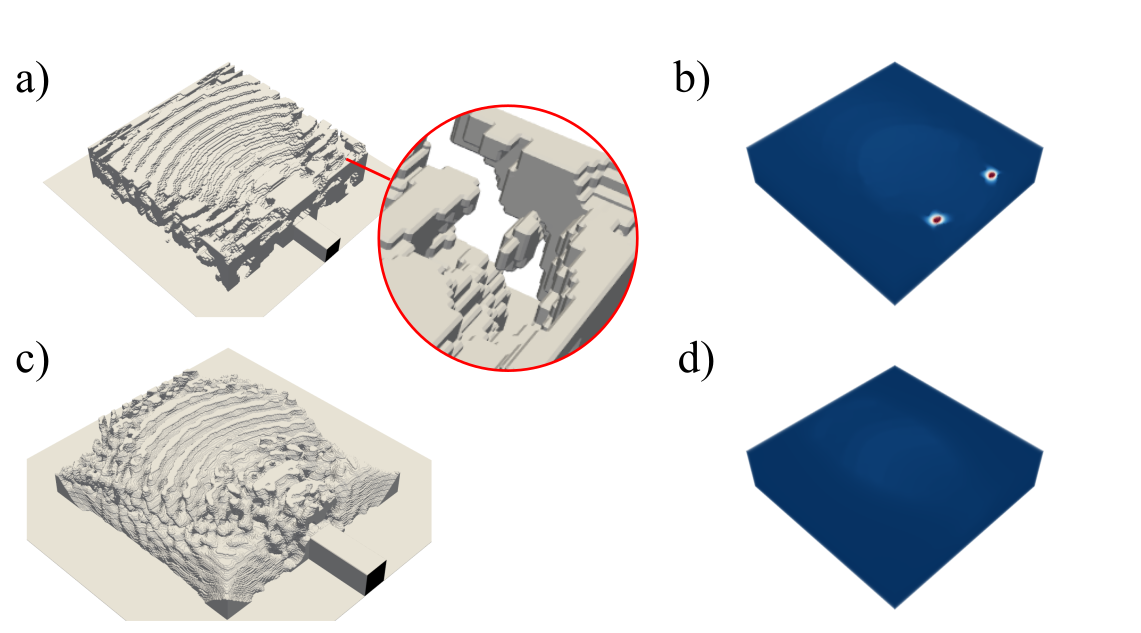}
    \caption{a) Optimized design without the heat solver to enforce connectivity, the inset shows the disconnected, floating part. b) Heat distribution of the heat generated by the material in a). The non-connected parts are clearly visible as hotspots. c) Optimized design with the heat solver. d) Heat distribution of the heat generated by the material in c). Both heat plots are normalized with respect to each other.}
    \label{fig:no-heat}
\end{figure}

Without the heat solver, the optimization finds a design that performs slightly better with up to $84\%$ coupling efficiency. But the lack of a heat solver results in disconnected, floating parts at the front of the design, which appear as hotspots in the heat distribution (see Fig.~\ref{fig:no-heat}). We only show the heat generated by the material, as both structures do not have any isolated pockets of void.
Even though the designs differ, adding the heat solver decreased the coupling efficiency by only $2\%$, indicating that our choice of threshold values, $\tau$, is sufficient to promote connectivity with minimal impact on performance.
Another factor contributing to the low coupling efficiency is that a grating coupler design naturally tends to converge into connected structures. This is also why most of the time, the non-connected structures are small, floating artifacts.

Since we did not optimize for broadband behavior, the coupling efficiency drops off away from the central wavelength.
Nonetheless, a coupling efficiency of above $80\%$ can still be achieved over a wavelength range of $\SI{20}{nm}$ for the $x$-polarized wave and $\SI{40}{nm}$ for the $y$-polarized wave, which puts it into a comparable broadband range as similar, two-dimensional, polarization sensitive grating couplers \cite{huang_compact_2025}.
Fig.~\ref{fig:eps_field} shows cross-sections of the device and the normalized field distribution for each polarization.
Looking at the cross-section in Fig.~\ref{fig:eps_field} a), the design superficially resembles a two-dimensional, density-based topology-optimized grating coupler.
However, in contrast to the two-dimensional variants, the three-dimensional design has a slanted grating profile (see Fig.~\ref{fig:sketch_and_design} b) and Fig.~\ref{fig:eps_field} b)), which contributes to its enhanced optical performance by providing more favorable scattering angles and mode shaping.
The slanted grating itself seems to emulate a blazed grating coupler \cite{guo_blazed_2023}, while also facilitating the coupling of both polarizations into their respective modes.
Additionally, by placing a mode source in the waveguide and propagating the wave into the grating coupler, we can also calculate the directionality $D = P_\text{up} / (P_\text{up} + P_\text{down})$. Averaged over the TE$_{00}$ and TM$_{00}$ modes, the design reaches a high directionality of $D=95\%$.

Additional geometric features emerge from the heat-based regularization. Since the heat objective favors highly thermally conductive structures, the optimization tends to introduce bulkier structures, particularly in regions that are not critical to the optical performance. Despite using a softplus activation to relax the constraint once connectivity is reached, these effects remain visible in the final design.
One such example is that the grating structure does not extend to the full width of the design region compared to the design optimized without the heat solver. Towards the edge of the design region, pure material or pure void emerges, as these dissipate heat into the heat sinks more effectively without significantly affecting the electromagnetic performance.
Another interesting feature that emerges is that the grating itself is suspended above the substrate in a bridge-like manner with a large air gap.
This seems to be partially due to the low refractive index contrast between the design and the substrate.
Interestingly, the design optimized without the heat solver also suspends the grating, but uses pillar-like structures underneath the grating instead.
By introducing a layer of air between the substrate and the material, a higher refractive index contrast can be achieved, improving the electromagnetic performance of the design while minimally impacting the heat performance, as the grating can be suspended at the bulkier edges.

\begin{figure}
    \centering
    \includegraphics[width=\linewidth]{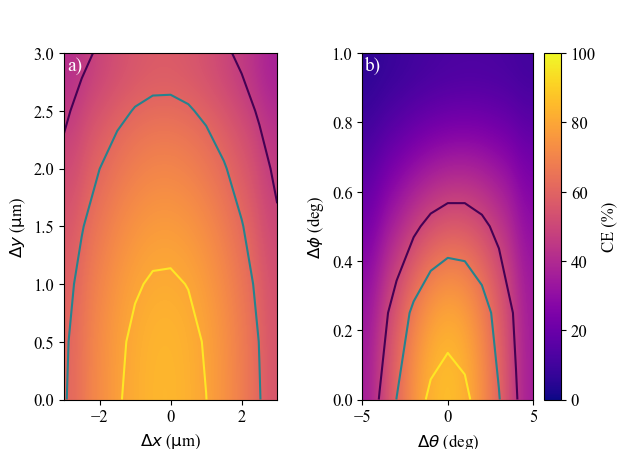}
    \caption{Coupling efficiency of the $\qtyproduct{18 x 18 x 4.5}{\upmu m}$ device with regard to lateral misalignment a) and angular misalignment b) of the fiber. Each isoline marks a subsequent loss of the coupling efficiency by $\SI{1}{dB}$. Note that the data shown here is interpolated.}
    \label{fig:variations}
\end{figure}

To test the robustness of our design against misalignment of the fiber, we analyze typical variations that can occur in an experimental setup. First, the coupling efficiency if the fiber position is shifted by $\Delta x \in [\SI{-3}{\upmu m}, \SI{3}{\upmu m}]$ in $x$-direction and $\Delta y \in [\SI{0}{\upmu m}, \SI{3}{\upmu m}]$ in $y$-direction. Second, we analyze the coupling efficiency for angular misalignment $\Delta \theta \in [-5^\circ, 5^\circ]$ and $\Delta \phi \in [0^\circ, 1^\circ]$. Here, $\theta$ represents the polar angle and $\phi$ the azimuth angle of the propagation axis in the plane orthogonal to the injection axis. In practice, $\theta$ represents the illumination angle we have set, and $\phi$ will rotate the angle of illumination away from the center. We note that changing $\phi$ will also rotate the polarization angle.
Both results can be seen in Fig.~\ref{fig:variations}. For the spatial and angular misalignment, we also indicate the locations of the insertion losses of $\SI{1}{dB}, \SI{2}{dB}$, and $\SI{3}{dB}$.
We can see that up to roughly $\SI{1}{\upmu m}$ of misalignment in the spatial direction still leads to sub-dB device performance. Overall, the design seems fairly robust to spatial misalignment, given that the optimization did not account for it.
As expected, the angular misalignment is less robust than the spatial misalignment. The design is especially vulnerable to misalignment in the azimuth angle, as this also changes the angle of polarization. Even a misalignment of $\Delta \phi = 0.1^\circ$ will degrade the performance by more than $20\%$. 

Robust topology optimization techniques can be employed to improve the robustness of the structure at the cost of decreasing the overall performance of the device \cite{sigmund_manufacturing_2009}.
One method of increasing robustness against spatial and angular misalignment would use a figure of merit which averages over an ensemble of simulations, each with a different randomly oriented fiber \cite{grieshammer_continuous_2024}. 

It is also possible to increase the bandwidth of the structure. Two common approaches include averaging over the desired bandwidth rather than optimizing a single wavelength, or choosing the wavelength with the worst coupling efficiency at each step, steadily improving the lower bound of the bandwidth range.

As for robustness against fabrication deviations, 3D nanoprinting tends to be sensitive to variations in the polymerization threshold, leading to a slightly eroded or dilated printed structure.
Erosion and dilation can also be addressed by robust topology optimization. By simulating two additional designs at different binarization thresholds $\alpha$ and averaging over all three designs at every iteration during the optimization, a design can be found that is less prone to erosion and dilation deviations.
Statistical deviations should also be taken into account in a similar manner as designing the device to be robust against misalignment.
One method would be to add and subtract Gaussian noise at every point to emulate local erosion and dilation. This will then lead to designs that are more robust against statistical imperfections during fabrication \cite{elbek_tailoring_nodate}.
It should be noted that statistical imperfections are less of an issue in 3D nanoprinting than in SOI designs, as the length scale at which they occur tends to be very small in 3D nanoprinting.

Practically speaking, the design and its applicability are limited by the 3D nanoprinting technology. Namely, due to the minimum feature size which can be printed and the printing speed, which determines the throughput and scalability of fabrication \cite{Hahn:19}.
Our minimum feature size is at the lower end of what is fabricable with 3D nanoprinting techniques. Increasing the feature size naturally degrades the coupling efficiency. We show a $\qtyproduct{20 x 20 x 5}{\upmu m}$ with a minimum feature size of roughly $\SI{500}{nm}$ which can reach a polarization-averaged coupling efficiency of $75\%$. The design is shown in the supplementary material (Fig.~\ref{fig:500nm}).

\section{Discussion}
In this work, we present a polarization-insensitive grating coupler, which achieves a coupling efficiency of more than 80\% for both polarizations, at a target wavelength of $\SI{1.55}{\upmu m}$. 
Our design process considers 3D nanoprinting as the fabrication method, allowing us to access all three dimensions in our design space and thus reach higher efficiencies while also staying polarization insensitive.
By using density-based topology optimization, we can efficiently optimize a design with more than $3$ million degrees of freedom. 
We use the virtual heat strategy to promote connectivity in both the material and the void, in addition to filtering and thresholding, to ensure that our designs can be fabricated with 3D nanoprinting.

While 3D nanoprinting cannot compete with the sheer throughput of CMOS-compared lithography, it is a promising, cost-efficient alternative for smaller and medium-sized companies to manufacture PICs.
Not only does 3D nanoprinting offer a flexible technology, enabling the printing of almost any geometry on demand, but it also opens the third dimension for design and optimization.
As shown in this work, access to another dimension in the design space paves the way for highly efficient nanophotonic designs and optimization techniques, which can overcome the bottlenecks of planar designs.

\section*{Acknowledgment}
The authors gratefully acknowledge the computing time provided on the high-performance computer HoreKa by the National High-Performance Computing Center at KIT (NHR@KIT). This center is jointly supported by the Federal Ministry of Education and Research and the Ministry of Science, Research and the Arts of Baden-Württemberg, as part of the National High-Performance Computing (NHR) joint funding program (https://www.nhr-verein.de/en/our-partners). HoreKa is partly funded by the German Research Foundation (DFG).

We thank Flexcompute for providing a licence for Tidy3D, that was used for the electromagnetic simulations in this work.

\section*{Data Availability}
The code to reproduce the datasets analyzed during the current study are available in the GitHub repository,\newline
\url{https://github.com/OlloKuster/3D-Grating_Coupler}.

\bibliographystyle{IEEEtran}
\bibliography{pol_ind_gc}
\newpage

\begin{wrapfigure}{l}{25mm} 
    \includegraphics[width=1in,height=1.25in,clip,keepaspectratio]{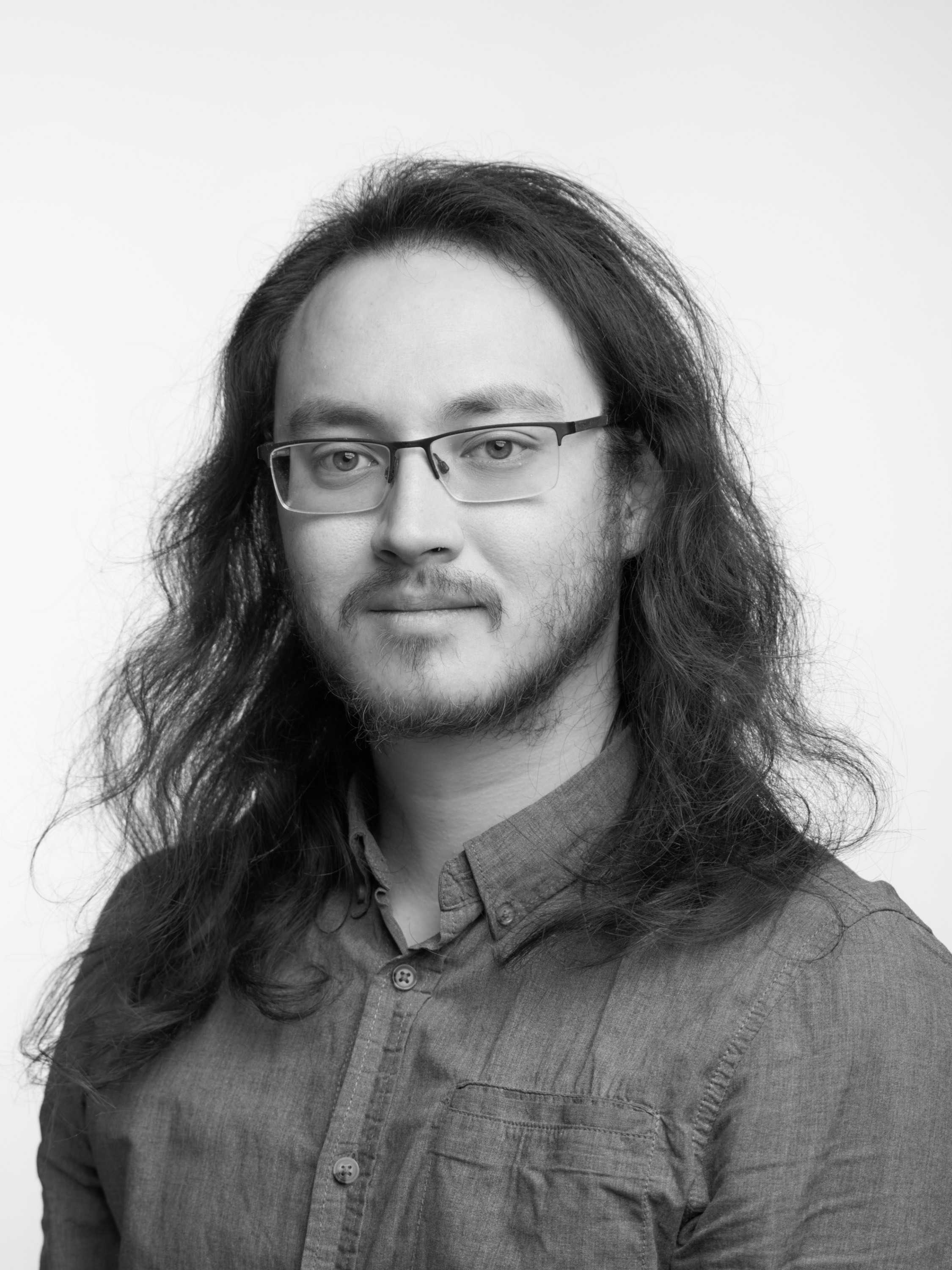}
\end{wrapfigure}\par
\textbf{Oliver Kuster} received his Master's degree in physics at the Karlsruhe Institute of Technology, Karlsruhe, Germany, in 2023.
He is currently working as a Ph.D. student under the supervision of Carsten Rockstuhl. His research interests include inverse design for nanophotonic applications, with a special focus on designs for 3D nanoprinting and computational nanophotonics.\\

\textbf{Yannick Augenstein} received the Ph.D. degree in physics from the Karlsruhe Institute of Technology, Karlsruhe, Germany, in 2024, under the supervision of Prof. Carsten Rockstuhl. He is currently a Computational Scientist with Flexcompute, where he focuses on developing tools for inverse design for photonics and electromagnetics. His research interests include inverse design, computational nanophotonics, and high-performance computing.\\

\begin{wrapfigure}{l}{25mm} 
    \includegraphics[width=1in,height=1.25in,clip,keepaspectratio]{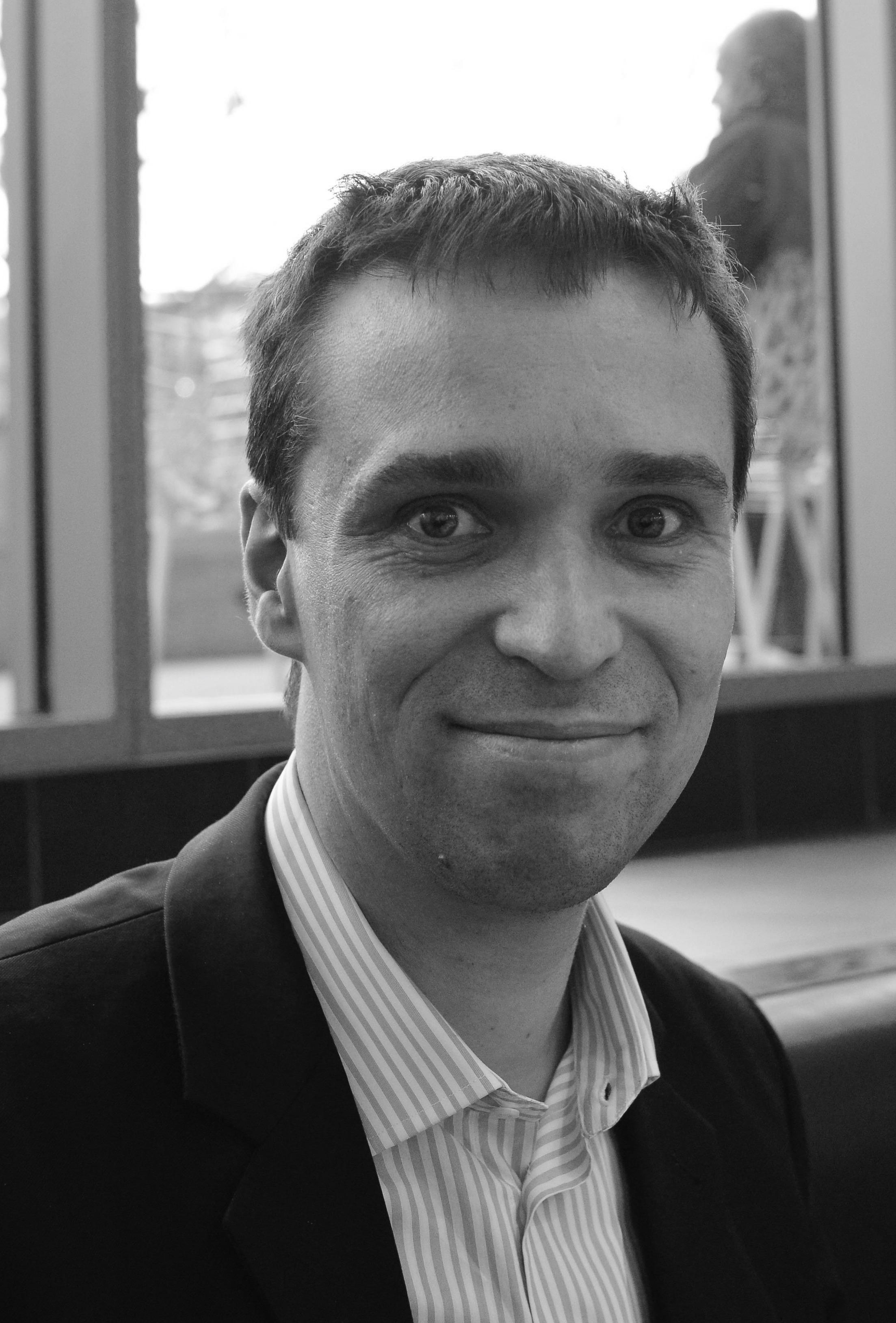}
\end{wrapfigure}\par
\textbf{Carsten Rockstuhl} received a Ph.D. from the University of Neuch\^{a}tel, Neuch\^{a}tel, Switzerland 2004. After a PostDoc period at AIST in Tsukuba, Japan, he has been since 2005 with the Friedrich Schiller University of Jena, Jena, Germany. In 2013, he was appointed a full professor at the Karlsruhe Institute of Technology, Karlsruhe, Germany. He works on many aspects in the context of theoretical and computational nano-optics. He serves the community as an editor with multiple journals. Moreover, he is a member of the Karlsruhe School of Optics \& Photonics, where he currently acts as the dean of study, the Max Planck School of Photonics, and is a fellow of Optica.\\

\begin{wrapfigure}{l}{25mm} 
    \includegraphics[width=1in,height=1.25in,clip,keepaspectratio]{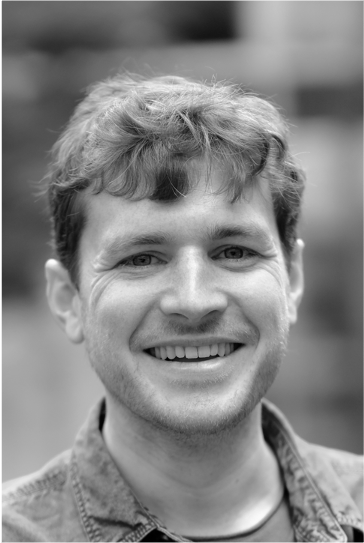}
\end{wrapfigure}\par
\textbf{Thomas Jebb Sturges} received a Ph.D. in theoretical physics in 2017 from the University of Exeter, U.K. From 2017 to 2022, he was a Postdoctoral Researcher with the University of Warsaw, Poland. From 2022 to 2025, he was an Alexander von Humboldt Research Fellow at the Karlsruhe Institute of Technology, Germany. He continues to work there. He has worked on a broad range of topics within computational photonics, theoretical quantum optics, topological photonics, strong light-matter coupling, and numerical modelling of polaritons. His current research focuses on photonic inverse design across a range of applications.

\section{Supplementary Material: The $\qtyproduct{20 x 20 x 5}{\upmu m}$ design}
\begin{figure}
    \centering
    \includegraphics[width=\linewidth]{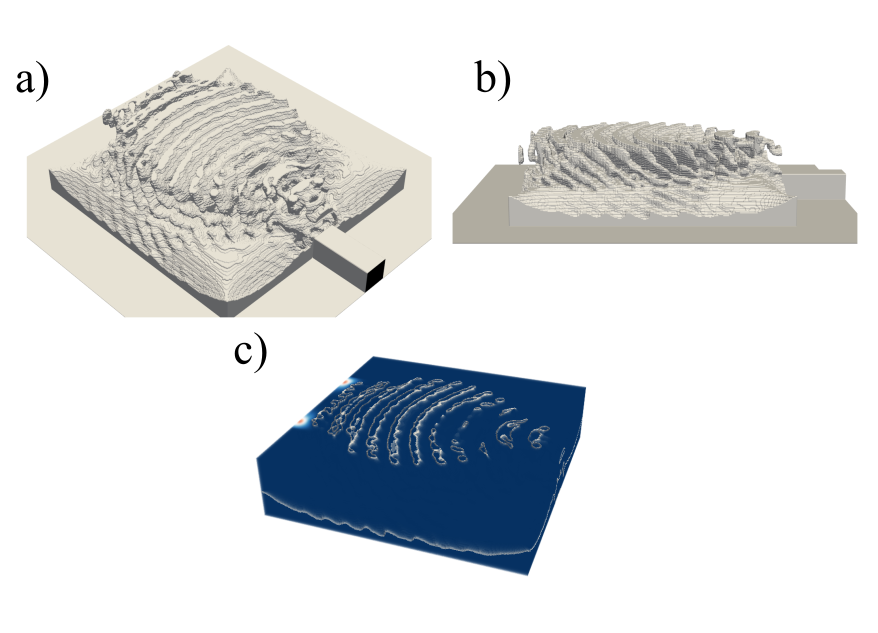}
    \caption{a) + b) The $\qtyproduct{20 x 20 x 5}{\upmu m}$ grating coupler design from different angles. c) The heat distribution of the material of this design, with the non-connected parts showing up as hotspots.}
    \label{fig:20}
\end{figure}

To compare the tradeoff between efficiency and footprint, the $\qtyproduct{20 x 20 x 5}{\upmu m}$ grating coupler is shown in Fig.~\ref{fig:20}. While it does appear to be similar to the $\qtyproduct{18 x 18 x 4.5}{\upmu m}$ design, floating elements appear during the optimization. These floating elements appear as hotspots in the heat distribution.
The $\qtyproduct{20 x 20 x 5}{\upmu m}$ grating coupler design can reach $85\%$ coupling efficiency, but increasing the connectivity constraints will lower the coupling efficiency.
Alternatively, the problem can be parametrized by applying the heat solver on the filtered and binarized density, but this also led to a degradation of the coupling efficiency and still does not guarantee the connectivity.

\section{Supplementary Material: Minimum feature size of $\SI{500}{nm}$}
\begin{figure}[!h]
    \centering
    \includegraphics[width=\linewidth]{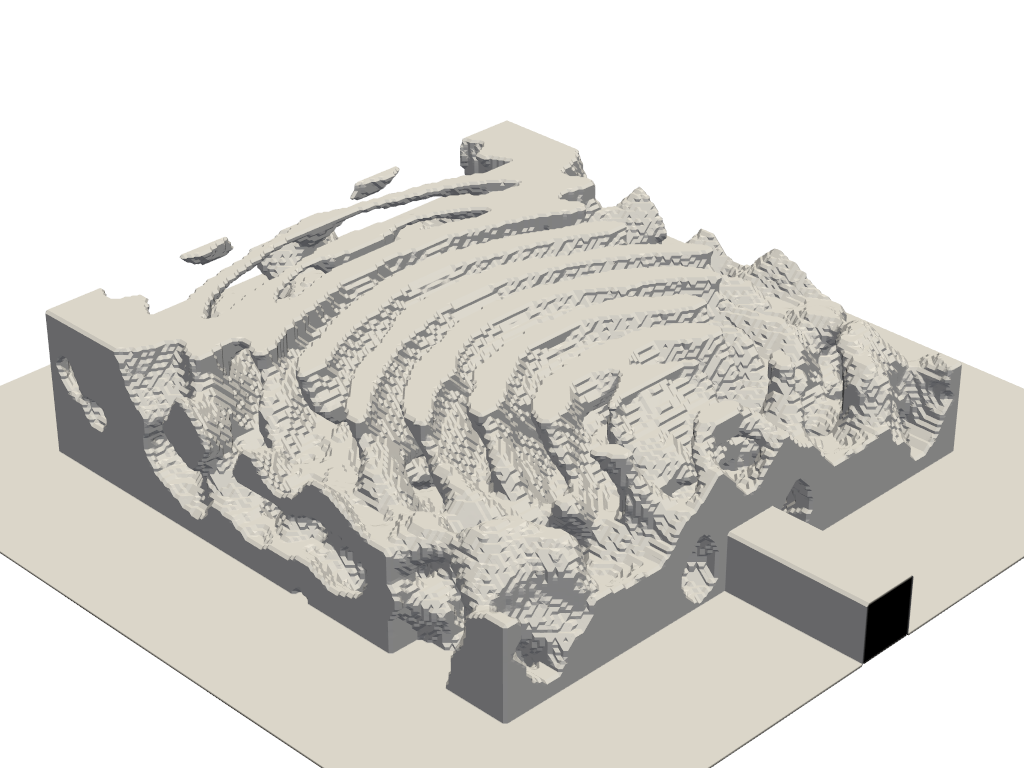}
    \caption{A $\qtyproduct{20 x 20 x 5}{\upmu m}$ grating coupler design with a minimum feature size of roughly $\SI{500}{nm}$.}
    \label{fig:500nm}
\end{figure}

We show a $\qtyproduct{20 x 20 x 5}{\upmu m}$ grating coupler design in Fig.~\ref{fig:500nm}. Note that the structure also has non-connected parts at its back, despite the heat solver.
Due to the larger feature size, the coupling efficiency has decreased to roughly $75\%$ compared to its counterpart with the lower minimum feature size. 

\begin{figure*}
    \centering
    \includegraphics[width=\linewidth]{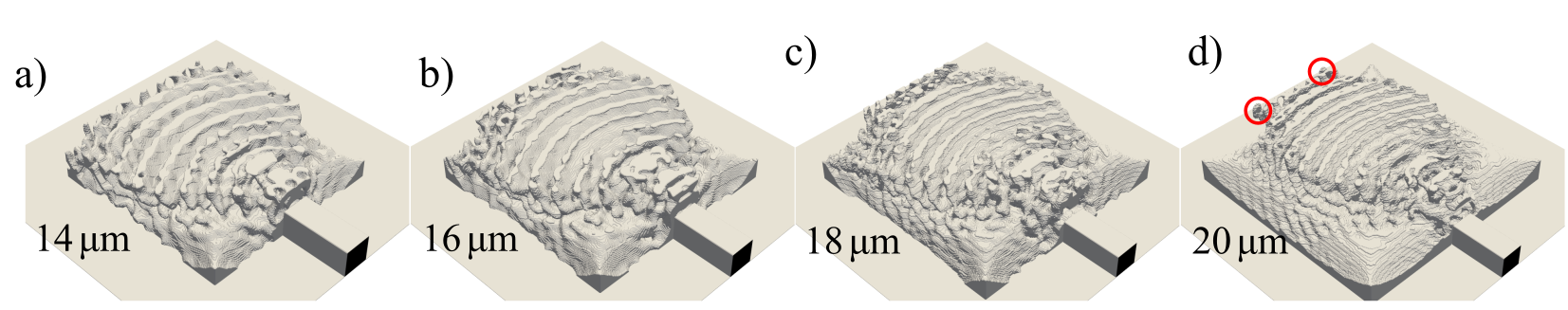}
    \caption{Four different designs with different base width. a) A $\qtyproduct{14 x 14 x 3.5}{\upmu m}$ polarization independent grating coupler. b) A $\qtyproduct{16 x 16 x 4}{\upmu m}$ polarization independent grating coupler. c) A $\qtyproduct{18 x 18 x 4.5}{\upmu m}$ polarization independent grating coupler. The $\qtyproduct{18 x 18 x 4.5}{\upmu m}$ is the best performing, still completely connected polarization independent grating coupler. d) A $\qtyproduct{20 x 20 x 5}{\upmu m}$ polarization independent grating coupler. The $\qtyproduct{20 x 20 x 5}{\upmu m}$ polarization-independent grating coupler starts to exhibit free-floating parts, which are highlighted inside the red circles.}
    \label{fig:all-four}
\end{figure*}

\end{document}